# Target domains in nanometric Permalloy disks with columnar structure

Svetlana Ponomareva[1], Robert Morel[1], Hélène Joisten[1,2], Philippe Sabon[1] and Bernard Dieny[1]

[1]Univ. Grenoble Alpes, CEA, CNRS, Spintec, 38000 Grenoble, France
[2]Univ. Grenoble Alpes, CEA, LETI, 38000 Grenoble, France
Contact e-mail: robert.morel@cea.fr

**Abstract**
We conducted a thorough experimental and numerical study of the micromagnetic properties of Permalloy ($Ni_{80}Fe_{20}$) microdisks exhibiting target domain structures at remanence. Vortex configurations are quite common in such microdisks and correspond to an in-plane flux closure configuration of cylindrical symmetry with an out-of-plane magnetized core. In contrast, target domain configuration are observed in thicker microdisks and are characterized by a vortex configuration of the in-plane component of the magnetization superposed to an out-of-plane component of magnetization which oscillates as a function of the distance to the microdisk center resulting in the formation of concentric domains. The ratio of the out-of-plane oscillatory component of the magnetization to the in-plane vortex one increases with the thickness of the microdisk. Hysteresis loops were measured under in-plane and out-of-plane field. The results at remanence and under magnetic field could be interpreted by micromagnetic simulations in which the microdisks were described as an assembly of partially coupled columns representing the granular nanostructure of the films from which the microdisks were patterned. Quite original magnetization processes take place in these microdisks exhibiting target domain remanent configuration. These include in particular entire flipping of the domain configuration and annihilation/creation of ring domains.

1. **Introduction**

In magnetism, the micromagnetic configuration of a ferromagnetic sample results from the balance between different energy terms, in particular the magnetic anisotropy which provides a preferential orientation to the magnetization, the exchange energy which favors uniform magnetization and the magnetostatic energy which tends to break the magnetization into domains. A very large variety of domain structures are observed depending on the intrinsic properties of the ferromagnetic material (anisotropy, exchange, spontaneous magnetization) and on its shape. Of particular interest are domains presenting a cylindrical symmetry which appear in thin films or patterned cylindrical micro/nanostructures. As examples, magnetic bubble domains have been extensively studied in micron-thick rare-earth orthoferrites films where they were considered for magnetic recording applications [1]. A theory linking the static stability of these structures to both the material properties and sample geometry was developed, highlighting the requirement that the uniaxial anisotropy must overcome the stray field energy in order to stabilize the bubble [2].

A second example of magnetic structure with cylindrical symmetry is the vortex structure encountered in sub-micron thin disks of soft magnetic material [3]. In this case the tradeoff between the minimization of the magnetostatic energy and cost in exchange energy results in the formation of an in-plane circular vortex, characterized by an in-plane flux-closed magnetic configuration of cylindrical symmetry and an out-of-plane magnetized core at its center whose diameter is of the order of the exchange length $l_{exch} = \sqrt{2A/\mu_0 M_s^2}$, typically a few nanometers [4]. In this expression, $A$ is the exchange stiffness constant and $M_s$ the spontaneous magnetization, $\mu_0$ the vacuum permeability.

In some other cases the magnetic structure becomes more complex, as for instance in the cylindrical concentric ring domains structure (so-called target domain configuration) that have been observed in a variety of magnetic thin films patterned in the form of cylindrical dots, characterized by multiple nested concentric magnetic rings. Such micromagnetic configurations have been reported in systems with out-of-plane magnetization such as high magnetic anisotropy Co and FePt films [5,6] or perpendicular anisotropy multilayers [7,8] which are known to form labyrinth type of domain structures in continuous thin film form. However, this type of target domains have also been observed in low anisotropy NiFe and Ni films [9,10] in which the magnetization is essentially in-plane with a weak out-of-plane anisotropy component. The patterned elements thickness is generally of a few tens of nanometers, with lateral size in the 100 to 1000 nm range. This type of films combining easy-plane anisotropy with a slight out-of-plane anisotropy component, tends to exhibit an homogeneous in-plane magnetization at wafer level below a critical thickness $t_c$ and a so-called stripe domain structure above $t_c$ [11-17]. The critical thickness normalized by the exchange length was shown to be a function of the film quality factor $Q = \frac{K_u}{\frac{\mu_0}{2}M_s^2}$, where $K_u$ is the perpendicular anisotropy constant. In the limit of small $Q$ ($Q \ll 1$), $t_c \sim 2\pi\sqrt{A/K_u}$.

The origin of the perpendicular anisotropy in NiFe or CoFe based films was ascribed to stress induced during deposition in samples of composition departing from the zero magnetostriction composition (i.e. departing from $Ni_{81}Fe_{19}$ for NiFe alloys) [13-15], and/or to the columnar structure of sputtered samples [11,12], especially in samples deposited in poor vacuum in which impurities accumulate within the grain boundaries [11,12]. In some cases, the induced perpendicular has been shown to be non-uniform along the thickness of the film due to grain coarsening and evolution in the film texture [16]. The impact of the stripe domain formation on the microwave excitations has also been studied in continuous films [13,15] and patterned microstructures [17].

The target domain structure in thick magnetic Permalloy nanodisks which is the focus of the present paper can be seen as a combination of the vortex physics with the stripe domain one.

Several numerical simulation studies of the micromagnetic behavior of magnetic micro/nanodisks with a perpendicular anisotropy contribution have been undertaken to investigate these bubble or target magnetic configurations [6-8,10,18,19]. Most of these simulations assumed the presence of a strong perpendicular anisotropy. Numerical simulations with low anisotropy systems have received less attention [9,10]. They still require to consider a rather large perpendicular uniaxial anisotropy - attributed to built-in magnetostriction or resulting from a columnar grain structure [20] - to explain the formation of the bubble or targer domain structure.

More recently, magnetic skyrmions have attracted much attention in bulk non-centrosymmetric ferromagnets or ferromagnetic systems where they are stabilized under an applied magnetic field due to the Dzyaloshinskii–Moriya exchange interaction [21]. In the best case, the skyrmions size in these systems can be less than 10 nm. From a topological point a view, the bubble domains in thin films and the concentric domains in patterned dots represent a second class of skyrmions, with a much larger radius, where the stabilization results from the balance between the uniaxial anisotropy and the magnetostatic energy [20,22].

In this paper, we present a thorough experimental and numerical study on patterned micron-size Permalloy (Py = $Ni_{80}Fe_{20}$) disks exhibiting magnetic target domain configurations at remanence. We investigated the evolution of the magnetic structure as a function of the NiFe thickness. Considering that the composition of our film is very close to the zero magnetostriction composition of NiFe alloys (obtained for 81.7% of Ni) [12], we argue that the dominant source of perpendicular anisotropy in our samples is associated to their columnar structure as will be illustrated further. Therefore, we also conducted numerical simulations in which, instead of considering the presence of a significant uniform perpendicular uniaxial anisotropy induced by stress, we consider that the system exhibits a columnar magnetic grain structure with variable inter-grain coupling. The inter-grain coupling intensity allows for the modulation of the relative influence of the magnetostatic energy and exchange energy, giving rise to various magnetic configurations hereafter described.

## 2. Experimental

Arrays of cylindrical microholes were patterned by optical lithography in resist bilayers, with a negative ma-N resist on top of a positive PMMA (polymethyl methacrylate) resist bottom layer, followed by Py deposition using e-beam evaporation. With this process, the Py disk deposition in the microholes of the top ma-N resists takes place on the bottom PMMA resist surface, resulting in a granular texture (see Figure 1). A lift-off step was then performed to remove the Py deposit above the top ma-N resist thereby leaving an array of Py microdisks on the bottom PMMA resist [23]. In all the experiments, the disks diameter was fixed at 1.3 µm while the thickness was varied from 90 nm to 400 nm. The micromagnetic structure of disks in the as-deposited state and after application of in-plane and out-of-plane saturating magnetic fields was observed by magnetic force microscopy (MFM). In addition, in-plane and out-of-plane magnetization measurements were performed by vibrating sample magnetometer (VSM) at room temperature.

Figure 1(a) shows a scanning electron microscopy image of one of our NiFe disk. The relief of its top surface is characteristic of the columnar structure of such films deposited by sputtering or evaporation on amorphous or polycrystalline substrates [11]. The columns are also visible on the sidewalls of the disk. From the size of the domes emerging on top of the columns, we can estimate the column diameter to be of the order of 20 nm.

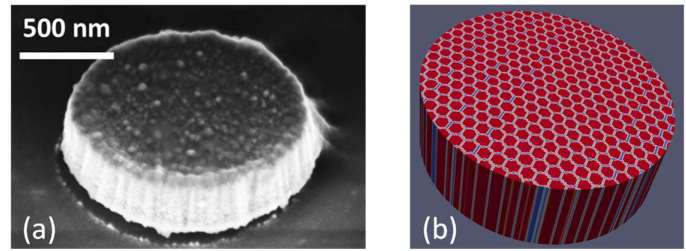

**Figure 1.** Disk morphology: (a) Scanning electron microscopy image of one of our patterned Permalloy disk illustrating the columnar nanostructure of the samples. (b) System used for the micromagnetic simulations consisting of a disk of 400 nm in diameter made of an hexagonal array of columns 20 nm in diameter.

Micromagnetic numerical simulations were carried out by solving the Landau-Lifshitz-Gilbert equation, using the Object Oriented MicroMagnetic Framework (OOMMF) finite difference simulation software [24]. The samples consist in disks with a diameter of 400 nm (smaller than the experimental diameter for the sake of computational time) and variable thickness between 60 nm and 300 nm. The magnetic energy comprises the exchange, the magnetostatic energy, the Zeeman energy whenever a field is applied but no contribution from any magnetocrystalline nor any source of uniaxial anisotropy.

The exchange and magnetization parameters used in the simulation are values typical for Permalloy: $A = 1.3 \times 10^{-11}$ J/m and $M_s = 800$ kA/m. With these values, the magnetostatic exchange length $l_{exch} = \sqrt{2A/\mu_0 M_s^2}$ is 5.7 nm. The cell size for the calculation is 2 nm x 2 nm x 10 nm. In order to emulate a columnar granular structure we consider that the disks consist in an array of packed hexagonal grains perpendicular to the disk plane, along the Z-direction (Fig.1b). The disk extends vertically from $Z = -t/2$ to $+t/2$ where $t$ is the disk thickness, $Z = 0$ representing the median plan of the disk. The diameter of the grains is approximately 20 nm. The boundary layer between grains has a thickness of one mesh cell. As a result, about 20% of the total number of mesh cells are located in grain boundaries. We further assume that the $A$ and $M_s$ values at the grain boundaries are reduced compared to their values in the bulk of the Py columns, with the same proportionality factor $\sigma$. This $\sigma$ parameter is the main parameter in this study. It characterizes the contrast of the magnetic properties ($A, M_s$) between the grain boundaries and the inner part of the grains.

For each thickness and $\sigma$ parameter values, the magnetic energy was minimized using, in average, five different magnetic configurations as initial conditions for the calculation. The initial configurations that have been used are the vortex configuration, target configurations with 2, 3 or 4 concentric domains, a fully Z-saturated configuration and a random configuration. It is observed that most of the time, the final configuration does not depend much on the initial configuration. Exceptions are however observed close to the critical thicknesses at which a change in the number of concentric domains occurs. In these situations, several remanent configurations may exist very close in energy.

## 3. Results and discussion

### 3.1. Experimental measurements

The remanent magnetization for disks in the as-deposited state are shown in Figure 2. Up to a thickness of 155 nm, most of the disks exhibit the common vortex structure [Fig. 2(a), 2(b)] [3]. Only a few exhibit a more complex metastable multi-domain in-plane structure. In a small thickness range, between 230 nm and 240 nm, the MFM imaging reveal a target structure with four concentric ring domains with alternating, up and down perpendicular component of the magnetization [Fig. 2(c), 2(e)]. Thicker disks, up to 400 nm, all exhibit a magnetic structure with three magnetic ring domains [Fig. 2(g), 2(h)]. In the thickness range where the target domain configurations are observed, the magnetic structure, for some of the disks, consists in spiraling domains or with an incomplete outer domain [Fig. 2(d), 2(f)].

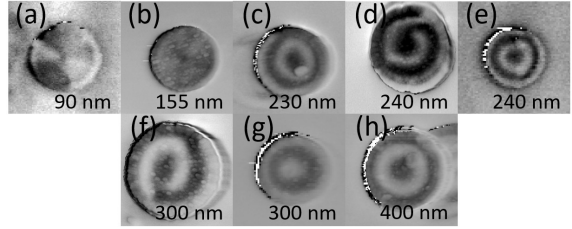

**Figure 2.** MFM images of Permalloy disks of 1.3 μm diameter, in the as-deposited state, and different thicknesses. (a, b) Vortex structure; (c, e) Four concentric ring domains structure; (d, f) Spiraling domains structure; (g, h) Three concentric ring domains structure.

Application of an in-plane (IP) saturating field results in a disordered remanent domain configuration, while application of an out-of-plane (OOP) saturating field always results in a remanent concentric target domain configuration.

The magnetization curves for arrays of vortex disks (thickness 155 nm) and target-domain disks (thickness 400 nm), with in-plane and out-of-plane applied field are shown in Figure 3. The IP magnetization for the vortex disks shows no remanence and an almost linear increase in magnetization up to the saturation field, here at 0.1 T [25]. We do not see the characteristic hysteresis associated with the annihilation/creation of the vortex when ramping the field up and down [3], which can be ascribed to the disk to disk variability over the ~$10^6$ microdisks comprised in the measured sample. The OOP magnetization is also linear at low field, with a higher saturation field, at 1 T, due to the easy-plane shape anisotropy. The magnetization curve for the target-domain disks with three concentric domains is similar in shape to the vortex one with again a very low remanence, but with a higher (resp. lower) low-field susceptibility in the IP (resp. OOP) configuration, and a less linear magnetization. It will be shown hereafter that with applied magnetic field, the target configuration evolves in a rather complex manner, with weak stripe domains (IP) and magnetization jumps associated with concentric domains nucleation/annihilation (OOP).

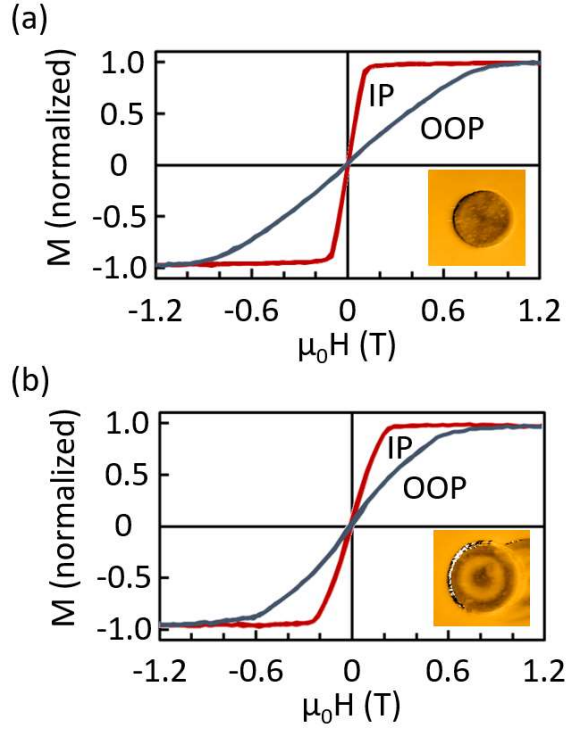

**Figure 3.** IP and OOP magnetization curve for 155 nm thick vortex disks (a) and 400 nm thick disks with target domains (b). Insets show the remanent magnetization state.

### 3.2. Micromagnetic simulations

In this section we discuss the magnetic nanostructure obtained with micromagnetic simulations, as a function of the thickness of the disk and the $\sigma$ coupling parameter between magnetic columns.

#### 3.2.1. Remanent magnetization and magnetic structure

The case with $\sigma = 1$ corresponds to homogeneous disks with no columns. Whatever the disk thickness, given the range of size and aspect ratio covered in the present study, the minimum energy configuration is the expected vortex magnetic structure [3,26]. At the opposite, with $\sigma = 0$, the grains are completely exchange-decoupled and the resulting structure is a disordered, metastable array of $\pm Z$-saturated magnetic columns, with no net magnetization for the disk.

With $\sigma$ values between 0 and 1 the magnetic structure shows a more complex behavior, with some examples shown in Figure 4. For the thinnest disks, the minimum energy configuration is again a vortex structure. Above a critical thickness $t_c$ which depends on the inter-column coupling parameter $\sigma$, it is observed that on top of the in-plane vortex component of the magnetization, part of the magnetization is pulled out of the disk plane, forming concentric ring domains with alternating up and down out-of-plane component of magnetizations. This evolution is described in more details below.

For low coupling ($\sigma = 0.05$), at a thickness of 60 nm, the micromagnetic structure essentially consists of a vortex with a narrow core of out-of-plane magnetization. A faint ring of blue contrast is visible around the core indicating a damped radial oscillation of the $M_z$ component of magnetization. As the thickness increases, the out-of-plane component of magnetization becomes more and more visible. A first well defined target domain structure appears at 90 nm, characterized by four concentric ring domains exhibiting a significant out-of-plane magnetization component alternatively up and down, on top of a still well defined in-plane vortex structure. Above 100 nm, the out-of-plane contrast reinforces and the magnetization switches to a configuration with three rings. Above 150 nm, the minimum energy configuration comprises two rings with fully saturated up and down magnetizations. The number of rings tends to decrease with the thickness of the film for a given intergrain coupling. This is reminiscent from the fact that because of the competition between magnetostatic and domain wall energies the stripe period in continuous films with out-of-plane anisotropy component tends to increase with film thickness in the investigated range of thickness [15,27].

For $\sigma = 0.1$, the four ring domains structure is not observed whatever the disk thickness. For the lowest thickness, the in-plane vortex structure dominates. The vortex core diameter increases with the disk thickness (compare the core width for 60 nm and 90 nm) and the radial oscillations of the $M_z$ component of magnetization becomes more visible. Above 120 nm, a clear target structure appears characterized by three rings. Above 150 nm, only two rings are observed.

For $\sigma = 0.2$, the trend of evolution of the domain structure versus disk thickness is similar to that observed for $\sigma = 0.1$. However for 120 nm, the out-of-plane contrast of the 3-rings target domain structure is less pronounced than for $\sigma = 0.1$ indicating a weaker out-of-plane component of magnetization and correlatively a stronger in-plane vortex component. As the thickness is increased, only two rings of alternating up and down magnetization are observed separated by the white ring associated with the in-plane vortex. The fact that for a given thickness, the number of rings tends to decrease as the intergrain coupling ($\sigma$) increases (for instance at 150 nm, 3 concentric domains are observed for $\sigma = 0.05$ while only two are observed for $\sigma = 0.2$) is ascribed to the increased cost in domain wall energy to create opposite domains as $\sigma$ increases. Correlatively, the domain wall width tends to increase with $\sigma$ which explains the quite visible white ring separating up and down domains observed for $\sigma = 0.2$.

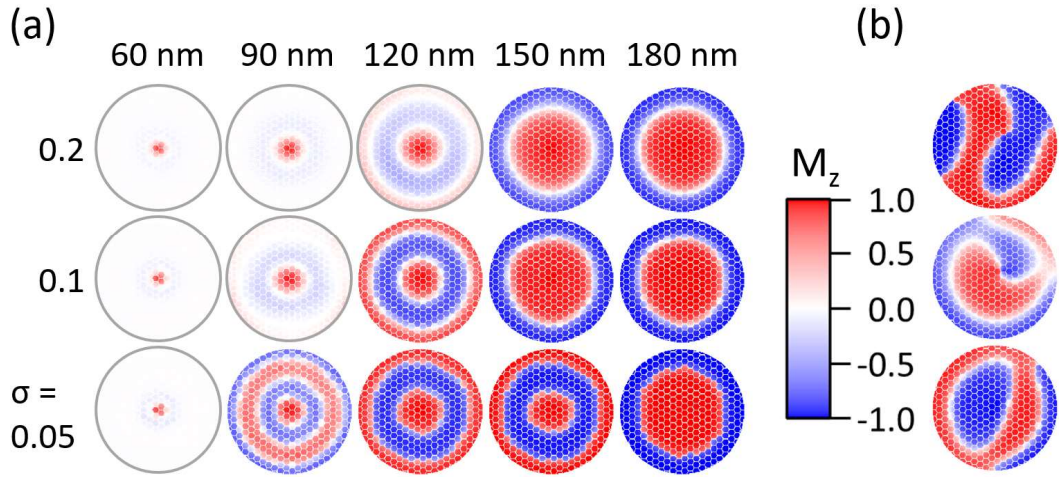

**Figure 4.** (a) Magnetic structure in the median plane of the disk (plane at half thickness), as a function of the thickness and $\sigma$ intergrain coupling parameter. (b) Examples of metastable configurations with disordered, spiral or incomplete ring domains, obtained with $\sigma = 0.1$ and 180 nm thickness (top image); $\sigma = 0.2$ and 150 nm (middle image); and $\sigma = 0.1$ and 150 nm (bottom image). The red/blue color scale refers to the normalized value of the $M_z$ magnetization component.

With $\sigma = 0.3$ and 0.5 (not shown) we observe that the magnetic structure evolves smoothly from a vortex configuration to a double rings domain structure above 150 nm. Finally, with $\sigma$ higher than 0.5, we observe only the vortex structure for all thicknesses up to 300 nm.

The numerical simulation indicates that the energy difference at the transition from one magnetic structure to the next is very small. This is particularly true with the structure with four rings that is very close in energy from the one with three concentric domains. In these situations, the energy minimization process is very dependent on the initial configuration and often results in metastable structures, as those illustrated in Figure 4(b).

Although the change in energy is small, a large amount of the magnetization that was in-plane in the vortex structure is rapidly pulled out-of-plane as the thickness of the disk increases but still remaining much below the disk diameter (diameter 400 nm). This trend towards perpendicular anisotropy is in fact strongly reinforced by the columnar structure of the sample. This is illustrated in Figure 5 where the average of the absolute OOP magnetization is plotted as a function of disk thickness for various intergrain coupling energies. Figure 5 shows that the out-of-plane component of the absolute magnetization increases with the disk thickness. This trend is all the more pronounced that the grains are more weakly coupled. For the disks with most weakly coupled grains ($\sigma < 0.2$), the out-of-plane component of the absolute magnetization rapidly reaches 80% of the saturation magnetization for disk thickness above 150 nm, indicating that the concentric domains are then almost saturated out-of-plane.

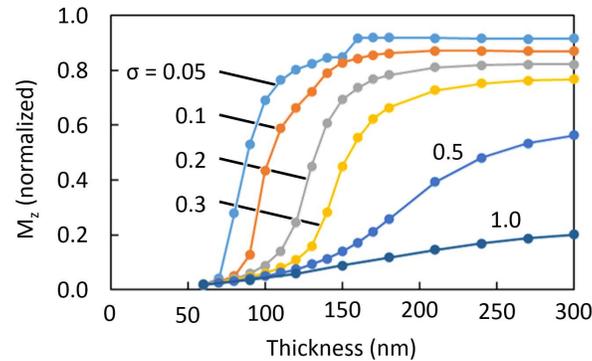

**Figure 5.** Average of the absolute value of the out-of-plane magnetization in the median plane of the disk ($Z = 0$) as a function of the disk thickness, for $\sigma = 0.05, 0.1, 0.2, 0.3, 0.5$ and $1.0$.

For disks exhibiting a target micromagnetic structure (thickness ranging between 60 nm and 300 nm), the $M_z$ profile within the median plan of the disk ($Z = 0$) versus position along the radius of the disk is shown in Figure 6. At 60 nm, the structure is a vortex with a 20 nm wide core and a very rapid decrease of the perpendicular magnetization component as one moves away from the disk center along a radius. For thicknesses larger than 90 nm, the magnetization evolves from a roughly sinusoidal profile (with four rings, $\sigma = 0.05$), to a steeper profile (with two rings) characterized by out-of-plane saturated concentric domains separated by a ~40 nm wide Bloch wall.

It is also observed that the positive and negative $Z$-magnetization for the concentric domains averaged on the whole disk volume nearly cancels out, leaving a net $Z$-magnetic moment always close to zero.

Moving away from the median plane of the disk, towards the top and bottom surfaces, the buckling of the magnetization of successive domains gives rise to magnetic torus with partial Néel-Bloch flux closure domains at both surfaces, as illustrated in Figure 7.

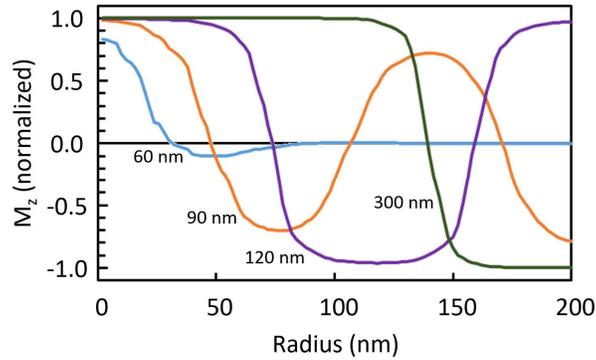

**Figure 6.** $M_Z$ magnetization versus position along the radius of the disk, within the median plan of the disk ($Z = 0$), for disks with $\sigma = 0.05$ of different thicknesses from 60 nm to 300 nm.

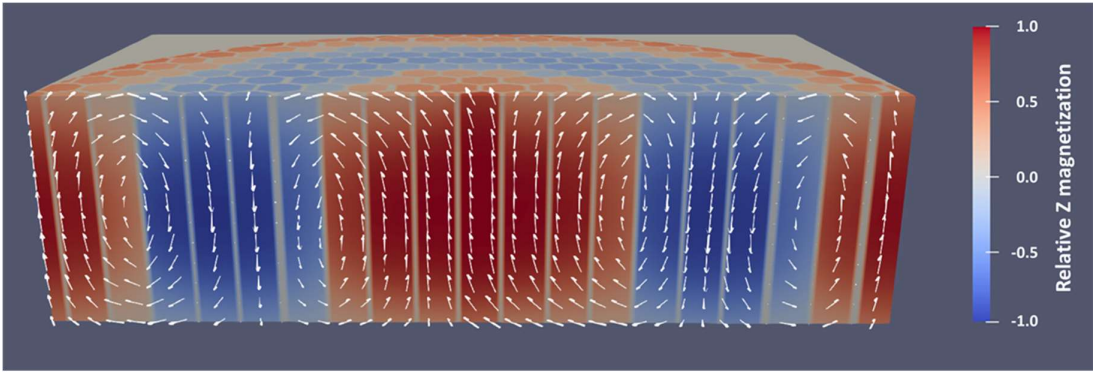

**Figure 7.** Cross sectional view of the magnetization in the bulk of a 120 nm thick disk with $\sigma = 0.05$, with three concentric domains. The arrows indicate the magnetization orientation within the cutaway $XZ$ plane. Color code indicates the normalized $M_z$ magnetization.

### 3.2.2. Magnetic structure with applied magnetic field

In addition to the zero field calculation, the evolution of the magnetic structure under magnetic field was calculated for IP and OOP applied magnetic fields, for disks exhibiting two- and three concentric domains at remanence.

The remanent state for both disks after application of an OOP saturating field is always with the outer domain magnetization in the direction of the applied field [Fig. 8].

If an increasing OOP magnetic field [Fig. 8(a)] is applied opposite to the outer domain magnetization, the outer ring domain shrinks and, at some point (between 100 mT and 125 mT for the two-domain target structure and between 50 mT and 75 mT for the three domain target structure), a flip in the magnetic structure occurs with the perpendicular magnetization of the outer domain lining up along the applied field, and a single inner circular domain with opposite magnetization appears. It is interesting to observe that the same two-domain structure is obtained with both the two-domain target and three-domain target initial configurations. Still increasing the field leads to the shrinking and gradual disappearance of the core domain, with saturation of the magnetization in the applied field direction.

When the OOP field is decreased from above the saturation field, the magnetization is homogeneous in the perpendicular direction, down to a nucleation field where the transition to the target-domain structure occurs.

If, again starting from the remanent configuration, the OOP applied field is applied parallel to the outer domain magnetization, this outer domain grows continuously at the expense of the core, until the latter shrinks and disappears at saturation.

If the field is applied in-plane, the micromagnetic configurations lose their cylindrical symmetry. Examples of magnetic structure with IP applied field are shown in Figure 8(b). At low fields, the center domain moves transversely to the field direction, as has commonly observed in vortex disks. At some point, the outer domain reverses on half of the disk perimeter and the magnetic structure transforms into a less-symmetrical pattern, with an incomplete outer domain that is, again, similar in both the two-domain and three-domain disks (although with opposite magnetization). At larger fields, this transforms suddenly into a weak stripe domains structure, with four stripes in the initially two-domain target and five in the initially three-domain target disks. The OOP magnetization in the stripe domains gradually vanishes when the IP field increases and the disk magnetization fully saturates in-plane. Under decreasing applied IP fields, starting from saturation, the same magnetic structures are observed. However, this time, the stripe domain structure persists over most of the field range down to a critical field at which the nucleation of the less-symmetrical domain structure takes place, rapidly followed by the recovery of the target-domain remanent structure.

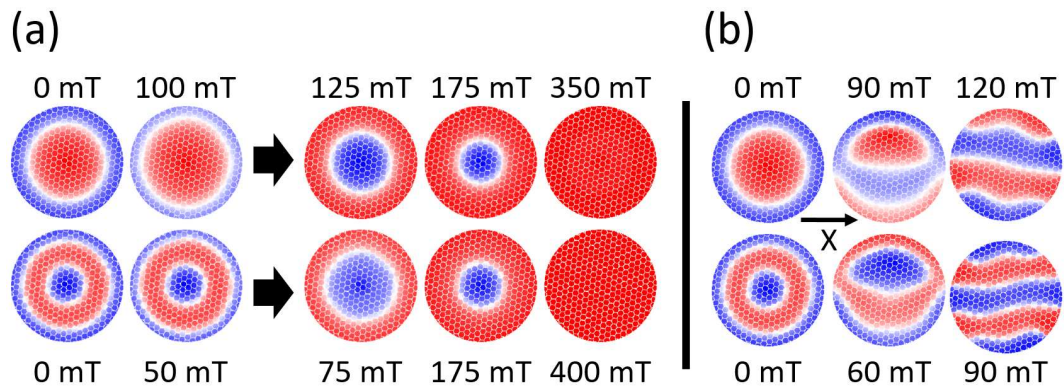

**Figure 8.** Magnetic structure for disks with applied fields. The remanent initial state was obtained after saturation in a negative OOP field. Top row: two-domain target disk ($\sigma = 0.2$, $t = 150$ nm). Bottom row: three-domain target disk ($\sigma = 0.1$, $t = 120$ nm). (a) Increasing OOP applied field. The arrows indicate a flip in the magnetic configuration, with the outer ring domain lining up along the applied field and, for the three-domain structure, the disappearance of one of the core domains. (b) Magnetic structure with IP field applied along the indicated X direction. Color code as in Figure 4.

### 3.2.3. Magnetization curve

Calculated magnetization curves for the two and three-target domain disks are shown in Figure 9. The shape of the curves is similar for the two types of disks, with slight variations in susceptibility and saturation fields. The hysteretic part in the IP curves corresponds to a difference in magnetization for the structures with incomplete outer domain (increasing field) and the stripe domain structure (decreasing field). The same feature is observed for the two types of disks. The low-field hysteresis in the OOP curves corresponds to the flip in the target-domain structure described in Figure 9(a).

Furthermore, the high field step in the field-decreasing OOP curves corresponds to the nucleation of the center zone in the target domain structure.

The calculated hysteresis loops can be compared with the measured OOP and IP curves of an array of 400 nm thick Py disk (diameter 1.3 $\mu$m) with a three target-domain configuration [Figure 3(b)]. The main difference is the absence of hysteresis in the experimental curves, which can be ascribed to a disk to disk distribution in the magnetic properties for the arrays of disks which were measured. Apart from this and a somewhat larger OOP and IP saturation fields in the measurement, the shape of the calculated hysteresis curves is in fair agreement with the experimental results. This is a good indication that our model, based on the assumption that the Py film consists of partially coupled magnetic columns, can correctly describe both the equilibrium magnetization configuration and its behavior under applied field.

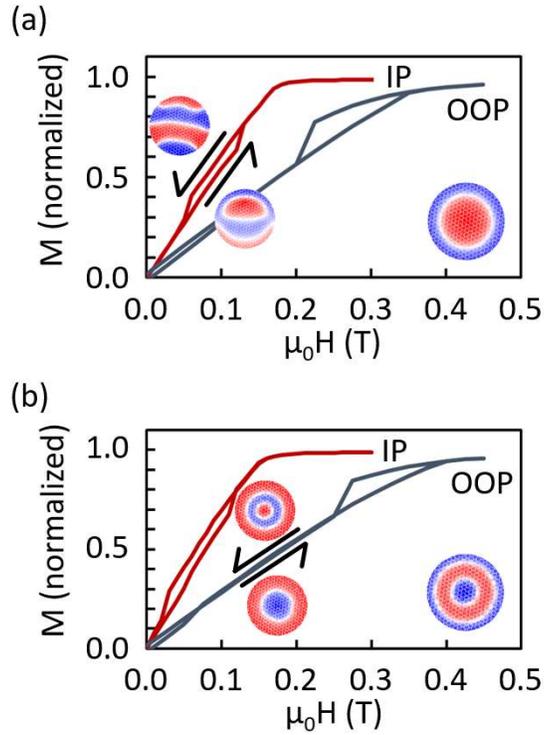

**Figure 9.** Calculated magnetization curves with IP and OOP applied field. (a) Two-domain target disk. (b) Three-domain target disk. Examples of difference in domain structure for increasing and decreasing fields are shown for IP (a) and OOP (b) curves. Lower right insets in each graph show the start magnetization state.

## 4. Discussion

The target-domain structure that is observed in Permalloy disks has similarities with the weak stripe domains structure observed in thin films. In the latter, the critical thickness $t_c$ for weak stripe nucleation, with no applied field, is given by $t_c = 2\pi\sqrt{A/K_u}$ [28]. Using this relation for our Permalloy disks, taking for $t_c$ the values given by the onset of the perpendicular oscillating magnetization reported in Figure 4, an effective anisotropy $K_u^*$ resulting from the columnar structure can be calculated. The $K_u^*$ values are found to increase as the coupling between columnar grains decreases, ranging from $3 \times 10^4$ J/m$^3$ (with $\sigma = 0.3$) up to $10 \times 10^4$ J/m$^3$ (with $\sigma = 0.05$). These values, although large, are similar to those previously measured in Ni thin films [10].

Interestingly, one can verify that there is an equivalence in terms of obtained micromagnetic structure between two models: i) the present numerical model considering the granular nanostructure of the film described by partially coupled vertical columns and ii) another model which would assume uniform exchange stiffness over the film and the presence of a uniform out-of-plane uniaxial anisotropy of amplitude $K_u^*$. This is illustrated by the results shown in Figure 10 where, by comparison with Figure 4, very similar structures (vortex and target-domain), with similar critical thicknesses as those obtained with the granular model, are observed.

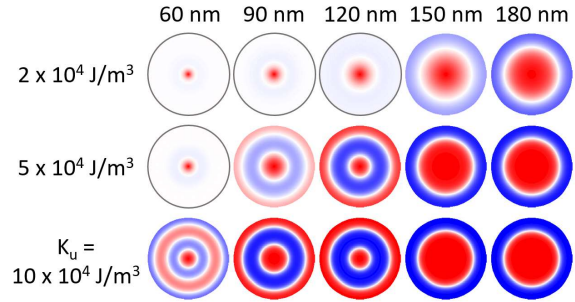

**Figure 10.** Magnetic structure over the disk median plan ($Z = 0$), as a function of disk thickness and uniform uniaxial perpendicular anisotropy $K_u$. Color code as in Figure 4.

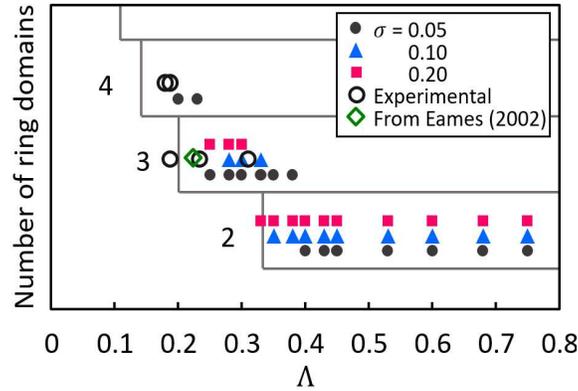

**Figure 11.** Number of ring domains as a function of the disk aspect ratio, for different values of the $\sigma$ coupling parameter. Also shown are the observed results for Permalloy disks, and a result from Eames (2002) [9]. Horizontal boxes show $\Lambda_n$, the lower limit for each number of ring domains (see text).

In addition to this, our results illustrate the fact that the number of ring domains ($n$) in a given disk (above the critical thickness) only depends on its aspect ratio $\Lambda = t/\phi$, with $\phi$ the diameter. This is illustrated in Figure 11 for the structures we have calculated, where the number of ring domains as a function of $\Lambda$ is given for different values for $\sigma$. Also shown in the figure are the results with Permalloy disks with 1.3 µm diameter [Fig. 2], and a result for Permalloy disks (with $t$ = 190 nm and $\phi$ = 850 nm) already reported in the litterature [9]. This shows that a given $n$ is only observed above a threshold value $\Lambda_n$, with a very good agreement between the calculations and the experiments. The exact micromagnetic calculation for $\Lambda_n$ is well beyond the scope of the present paper, but it is empirically determined that the relation $\Lambda_n = 1/(2n-1)$ agrees well with the results.

## 5. Conclusion

In conclusion, we have observed that micron-sized relatively thick Permalloy disks (thickness in the range 60 nm - 400 nm) can exhibit a variety of remanent micromagnetic structures from vortex magnetic structure to target domain structure consisting of several concentric ring domains. More complex metastable structures such as spiraling domains can also be observed.

In order to interpret the experimental observations, micromagnetic numerical simulations were carried out assuming that the film nanostructure consists in an assembly of columns having bulk Py properties in the inner part of the columns but variable magnetization and exchange-coupling parameter at the magnetic grain boundaries. In contrast to earlier models (even with soft magnetic materials), no intrinsic perpendicular anisotropy was considered in our model. Actually, an effective perpendicular anisotropy arises from the competition between each column out-of-plane shape anisotropy and the overall disk easy-plane shape anisotropy: Above a critical thickness, the resulting perpendicular anisotropy is strong enough to draw part of the magnetization out-of-plane, forming target domain configuration which are reminiscent of the weak stripe domains observed in thin films. We also observe that, above this critical thickness, the number of target domains is a unique function of the disk aspect ratio.

A more general conclusion is that the columnar nanostructure that we consider (a feature that is often present in, for instance, sputtered thin films) but often overlooked in modelling, can be at the origin of a significant perpendicular anisotropy contribution in soft magnetic materials.

**Acknowledgements**

This work has been supported by funding from the European Union's Horizon 2020 research and innovation program under grant agreement ABIOMATER No 665440.